\documentclass[12pt]{article}

\usepackage{graphicx,xspace,colortbl}

\usepackage{amsmath,amsthm,amsfonts}
\usepackage{fancybox}

\newtheorem{theorem}{Theorem}

\newtheorem{proposition}[theorem]{Proposition}
\newtheorem{corollary}[theorem]{Corollary}

\theoremstyle{definition}

\renewcommand{\|}{ { \lvert } }

\def\qed{\hfill$\sqcap\kern-8.0pt\hbox{$\sqcup$}$\\}
\def\beq{\begin{eqnarray}}
\def\eeq{\end{eqnarray}}
\def\beqq{\begin{eqnarray*}}
\def\eeqq{\end{eqnarray*}}

\def\be{\begin{equation}}
\def\ee{\end{equation}}

\newcommand{{\X}}{{\bf X}}
\newcommand{{\x}}{{\bf x}}
\newcommand{{\Z}}{{\bf Z}}
\newcommand{{\z}}{{\bf z}}
\newcommand{{\Y}}{{\bf Y}}
\newcommand{{\y}}{{\bf y}}
\newcommand{{\F}}{{\bf F}}
\newcommand{{\bbeta}}{{\bf \beta}}
\newcommand{{\bsigma}}{{\bf \sigma}}
\newcommand{{\bL}}{{\bf L}}
\newcommand{{\bW}}{{\bf W}}
\newcommand{{\bu}}{{\bf u}}
\newcommand{{\im}}{\mbox{Im}}

\title{A Fourier transform method for spread option pricing\thanks{Research supported by the
Natural Sciences and Engineering Research Council of Canada and
MITACS, Mathematics of Information Technology and Complex Systems
Canada}}

\author{T. R. Hurd\thanks{Dept. of Mathematics and Statistics, McMaster
University, Hamilton ON L8S 4K1, Canada ({\tt hurdt@mcmaster.ca}).}
    \     and Zhuowei Zhou\thanks{Dept. of Mathematics and Statistics, McMaster
University, Hamilton ON L8S 4K1, Canada.}}

\begin{document}

\maketitle

\begin{abstract}
Spread options are a fundamental class of derivative contract written on multiple assets, and are widely used in a range of financial markets. There is a long history of approximation methods for computing such products, but as yet there is no preferred approach that is accurate, efficient and flexible enough to apply in general models. The present paper introduces a new formula for general spread option pricing based on Fourier analysis of the spread option payoff function. Our detailed investigation proves the effectiveness of a fast Fourier transform implementation of this formula for the computation of prices. It is found to be easy to implement, stable, efficient and applicable in a wide variety of asset pricing models.
\end{abstract}



\section{Introduction}\label{introduction}
When $S_{jt}, j=1,2$ are two asset price procsses, the basic spread option with maturity $T$ and strike $K\ge 0$ is the contract that pays $(S_{1T}-S_{2T}-K)^+$ at time $T$. From the risk-neutral expectation formula, the time $0$ price of this option, assuming a constant interest rate $r$, must be
\begin{equation}
\mbox{Spr}(S_0;T, K)=e^{-rT}E_{S_0}[(S_{1T}-S_{2T}-K)^+]\ .\label{rnpricing}
\end{equation}
 
The literature on applications of spread options is extensive, and is reviewed by Carmona and Durrleman \cite{CarmDurr03} who explore further applications of spread options beyond the case of equities modelled by geometric Brownian motion, notably energy trading. For example, the ``crack spread'' is the difference between the prices of crude oil and natural gas. ``Spark spreads'' refer to differences between the price of electricity and the price of fuel: options on spark spreads are widely used by power plant operators to optimize their revenue streams. Energy pricing requires models with mean reversion and jumps very different from geometric Brownian motion, and pricing spread options in such situations can be challenging. 

Closed formulas for \eqref{rnpricing} are known only for a limited set of cases. In the Bachelier stock model, $S_t=(S_{1t},S_{2t})$ is an arithmetic Brownian motion, and in this case \eqref{rnpricing} has a Black-Scholes type formula for any $T,K$. In the special case $K=0$ when $S_t$ is geometric Brownian motion, \eqref{rnpricing} is given by the Margrabe formula \cite{Margrabe78}.

In the important case where $S_t$ is geometric Brownian motion and $K>0$, no explicit pricing formula is known. Therefore there is a long history of approximation methods. Numerical integration methods, typically Monte Carlo based, are often employed. Dempster and Hong \cite{DempHong00} introduced a numerical integration method based on double fast Fourier transforms  (FFT) that is efficient for geometric Brownian motion and certain more general asset price processes.  

Another stream of research develops analytical methods applicable to log normal models that involve linear approximations of the nonlinear exercise boundary. Such methods are often very fast, but their accuracy is usually not easy to determine.  Kirk \cite{Kirk95} presented an analytical approximation that performs well in practice.  Carmona and Durrleman \cite{CarmDurr03a} used a family of lower bounds for  the spread option price to obtain an analytical approximation formula that apparently leads to acceptable accuracy. Moreover these results extend to approximate values for the Greeks.  Deng, Li and Zhou \cite{DenLiZho06} have developed a systematic analytical approximation scheme which is perhaps even more reliable. 

Next to having analytical option pricing formulas, the fastest option pricing engines by numerical integration are usually those based on the fast Fourier transform methods introduced by Carr and Madan \cite{CarrMada00}. Their interest was in option pricing for geometric L\'evy process models like the variance gamma (VG) model, but their basic framework has been adapted to a host of other models where the characteristic function of the returns is known. 

The main purpose of the present paper is to give a new numerical integration method for computing spread options in two or higher dimensions via FFTs, applicable to any model for which the characteristic function of the joint return process is given analytically. Our method is completely different from prior approaches, and consequently has distinct advantages, notably the flexibility to include a wide range of desirable asset return models, with a competitive computational expense.

The results we describe all stem from the following fundamental new result that gives a Fourier representation of the basic spread option payoff function
 $P(x_1,x_2)=(e^{x_1}-e^{x_2}-1)^+$. Note that this spread option payoff is without loss of generality: We can reduce the general case $K\ne 0$ to $K=1$ by using scaling and interchange of $S_1$ and $S_2$.
 \begin{theorem} \label{mainprop}
 For any real numbers $\epsilon=(\epsilon_1,\epsilon_2)$ with $\epsilon_2>0$ and $\epsilon_1+\epsilon_2<-1$
 \begin{equation} 
P(x)=(2\pi)^{-2}\iint_{\mathbb{R}^2+i\epsilon} e^{iux'} \hat P(u) d^2u,\quad
 \hat P(u)=\frac{\Gamma(i(u_1+u_2)-1)\Gamma(-iu_2)}{\Gamma(iu_1+1)}\ .
\label{pfft}
 \end{equation}
 Here  $\Gamma(z)$ is the complex gamma function defined for $\Re e (z)>0$ by the integral 
 $\Gamma(z)=\int^\infty_0e^{-t}t^{z-1}dt$.
 We write $u=(u_1, u_2)$ and $x=(x_1,x_2)$ and the notation $x'$ denotes the (unconjugated) matrix transpose.
 \end{theorem}

Using this result, whose proof is given in the appendix, we will find we can follow the logic of Carr and Madan to derive numerical algorithms for efficient computation of a variety of spread options and their Greeks. The basic strategy to compute \eqref{rnpricing} is to combine \eqref{pfft} with an explicit formula for the characteristic function of the bivariate random variable $X_t=\log S_t$. For the remainder of this paper, we make a simplifying assumption that the dynamics of $X$ is {\it homogeneous}, which is equivalent to the following factorization of the characteristic function 
 \begin{equation}\label{charfn}
 E_{X_0}[e^{iuX_T'}]=e^{iuX'_0}\Phi(u;T),  \quad \Phi(u;T):=E_{0}[e^{iuX_T'}]\ .
 \end{equation}
Then, if $S_t=e^{X_t}$ is the price process for two traded assets, the spread option formula can be written
 \begin{eqnarray}
 \mbox{Spr}(S_0;T)&=&e^{-rT}E[(S_{1T}-S_{2T}-1)^+]\nonumber\\
 &=&(2\pi)^{-2}e^{-rT}\iint_{\mathbb{R}^2+i\epsilon} E_{X_0}[e^{iuX_T'}] \hat P(u) d^2u\nonumber\\
 &=&(2\pi)^{-2}e^{-rT}\iint_{\mathbb{R}^2+i\epsilon}e^{iuX'_0} \Phi(u;T) \hat P(u) d^2u\ .
\label{spread}
 \end{eqnarray}
 The Greeks are handled in exactly the same way. For example, the Delta $\Delta^1:=\partial \mbox{Spr}/\partial S_{10}$ is obtained as a function of $S_0$ by replacing $\Phi$ in \eqref{spread} by $\partial \Phi/\partial S_{10}$.
 
Explicit double Fourier integrals like this can be approximated numerically by a two dimensional FFT. Such approximations involve both a truncation and discretization of the integral, and the two properties that determine their accuracy are the decay of the integrand of \eqref{spread} in $u$-space, and the decay of the function $\mbox{Spr}$ in $x$-space. The remaining issue of computing the gamma function is not a real difficulty. Fast and accurate computation of the complex gamma function in for example, Matlab, is based on the Lanczos approximation popularized by \cite{Pressetal}\footnote{According to these authors, computing the gamma function becomes ``not much more difficult than other built-in functions that we take for granted, such as $\sin x$ or $e^x$''.}.

In this paper, we demonstrate how our method performs in three different models for spread options on two stocks, namely the geometric Brownian motion (GBM), a three factor stochastic volatility (SV) model and the variance gamma (VG) model. Section 2 provides the essential definitions of the three types of models, including explicit formulas for their bivariate characteristic functions.  Section 3 discusses how the two dimensional FFT can be implemented for our problem, and gives a heuristic picture of how the accuracy and speed will depend on the choices made in the implementation. Section 4 gives the detailed results of the performance of the method in the three models. In that section, the accuracy of each model is compared to benchmark values computed by an independent method for a reference set of option prices. We also demonstrate that the computation of the spread option Greeks in such models is equally feasible. Section 5 extends all the above results to spread options on three or more assets. While in such cases the formulation is simple, the resulting higher dimensional FFTs are in practice much slower to compute. 

\section{Three kinds of stock models}
\subsection{The case of geometric Brownian motion}
In the two-asset Black-Scholes model, the vector $  S_t=(S_{1t},S_{2t})$ has components
\[ S_{jt}=S_{j0}\exp[(r-\sigma_j^2/2)t+\sigma_j W^{j}_t], j=1,2\]
where $\sigma_1,\sigma_2>0$ and $W^{1},W^{2}$ are risk-neutral Brownian motions with constant correlation $\rho, |\rho|<1$. The joint characteristic function of $  X_T=(\log   S_{1T},\log S_{2T})$ as a function of $u=(u_1,u_2)$ is
of the form $e^{iuX'_0}\Phi(u;T)$ with
\begin{equation}\Phi(  u;T)=\exp[i  u( r T e- \sigma^2T/2)'-  u\Sigma  u'T/2]\ 
\end{equation}
where $  e=(1,1), \Sigma=[\sigma_1^2,\sigma_1\sigma_2\rho;\sigma_1\sigma_2\rho,\sigma_2^2]$ and $ \sigma^2=\mbox{diag}\Sigma$. Here we use implied matrix multiplication, and the $u'$ denotes the (unconjugated) matrix transpose.  Substituting this expression into \eqref{spread} yields the spread option formula
\begin{equation}\label{gbmspread} \mbox{Spr}(S_0;T)=(2\pi)^{-2}e^{-rT}\iint_{\mathbb{R}^2+i \epsilon}e^{iuX'_0}\exp[i  u(r   Te-  \sigma^2 T/2)'-  u\Sigma  u' T/2]\hat P( u)d^2u\ .
\end{equation}
As we discuss in Section 3, we recommend that this be computed numerically using the FFT. 

\subsection{Three Factor Stochastic Volatility Model}
The spread option problem in a three factor stochastic volatility model was given as an example by Dempster and Hong \cite{DempHong00}. Their model is defined by the SDE for $X=(\log S_1,\log S_2)$:
\begin{eqnarray*}
dX_1&=& [(r-\delta_1-\sigma_1^2/2 )dt+\sigma_1\sqrt{v}dW^1]\\
dX_2&=& [(r-\delta_2-\sigma_2^2/2 )dt+\sigma_2\sqrt{v}dW^2]\\
dv&=&\kappa(\mu-v)dt + \sigma_v \sqrt{v} dW^v
\end{eqnarray*}
where 
\begin{eqnarray*}
E[dW^1 dW^2 ] &=&\rho dt \\
EQ[dW^1 dW^v ] &=&\rho_1 dt \\
EQ[dW^2 dW^v ] &=& \rho_2 dt. 
\end{eqnarray*}
As discussed in that paper, it has the joint characteristic function $e^{iuX'_0}\Phi(u;T)$ where
\begin{eqnarray*}
\Phi(u;T)&=&\exp\Biggl[\left(\frac{2\zeta(1-e^{-\theta T})}{2\theta-(\theta-\gamma)(1-e^{-\theta})}\right) v(0)\\
&&+iu(r e-\delta)'T-\frac{\kappa\mu}{\sigma_v^2}\left[2\log\left(\frac{2\theta-(\theta-\gamma)(1-e^{-\theta T})}{2\theta}\right)+(\theta-\gamma)T\right]\Biggr]
\end{eqnarray*}
where
\begin{eqnarray*}
\zeta&:= &-\frac12\left[\left(\sigma_1^2u_1^2+\sigma_2^2u_2^2+2\rho\sigma_1\sigma_2u_1u_2\right)+i\left(\sigma_1^2u_1+\sigma_2^2u_2\right)\right]\\
\gamma&:=&\kappa-i(\rho_1\sigma_1u_1+\rho_2\sigma_2u_2)\sigma_\nu\\
\theta&:=&\sqrt{\gamma^2-2\sigma_v^2\zeta}\ .
\end{eqnarray*}

\subsection{Exponential L\'evy Models}

Many stock price models  are of the form $S_t= e^{ X_t}$ where $X_t$ is a L\'evy process for which the characteristic function is explicitly known. We illustrate with example of the VG process introduced by  \cite{MadaSene90}, the three parameter process $Y_t$ with L\'evy characteristic triple $(0,0,\nu)$ where the L\'evy measure is $\nu(x)=\lambda[e^{-a_+x}{\bf 1}_{x>0}+e^{a_-x}{\bf 1}_{x<0}]/|x|$ for positive constants $\lambda,a_\pm$. The characteristic function of $Y_t$ is
\begin{equation}
 \Phi_{Y_t}(u)=\left[1+i\left(\frac1{a_-}-\frac1{a_+}\right)u+\frac{u^2}{a_-a_+}\right]^{-\lambda t}.
 \end{equation}

To demonstrate the effects of correlation, we take a bivariate VG model driven by three independent VG processes $Y_1, Y_2, Y$ with common parameters $a_\pm$ and $\lambda^1=\lambda^2=(1-\alpha)\lambda,
\lambda^Y=\alpha\lambda $. The bivariate log return process $X_t=\log S_t$ is a mixture:
\begin{equation} X_{1t}=X_{10}+Y_{1t}+ Y_t;\quad X_{2t}=X_{20}+Y_{2t}+ Y_t\ .
\end{equation}
Here $\alpha\in[0,1]$ leads to dependence between the two return processes, but leaves their marginal laws unchanged. An easy calculation leads to the bivariate characteristic function $e^{iuX'_0}\Phi(u;T)$ with
\begin{eqnarray}
\Phi(u;T)&=&\left[1+i\left(\frac{1}{a_-}-\frac{1}{a_+} \right)(u_1+u_2)+\frac{(u_1+u_2)^2}{a_-a_+}\right]^{-\alpha\lambda t}\label{vgchar}\\&&\hspace{-1.7cm}\times
\left[1+i\left(\frac{1}{a_-}-\frac{1}{a_+} \right)u_1+\frac{u^2_1}{a_-a_+}\right]^{-(1-\alpha)\lambda t}\left[1+i\left(\frac{1}{a_-}-\frac{1}{a_+} \right)u_2+\frac{u^2_2}{a_-a_+}\right]^{-(1-\alpha)\lambda t}\ .\nonumber
\end{eqnarray}

\section{Numerical Integration by Fast Fourier Transform}
\label{nifft}
To compute \eqref{spread} in these models we approximate the double integral by a double sum over the lattice
\[\Gamma=\{  u(  k)=(u(k_1),u(k_2))|  k=(k_1,k_2)\in\{0,\dots, N-1\}^2\}
,\quad u(k)=-\bar u +k\eta\]
for appropriate choices of $N,\eta, \bar u:=N\eta/2.$
For the FFT it is convenient to take $N$ to be a power of $2$ and lattice spacing $\eta$ such that truncation of the  $u$-integrals to $[-\bar u,\bar u]$ and discretization leads to an acceptable error. Finally, we choose initial values $  X_0=\log  S_0$ to lie on the reciprocal lattice with spacing $\eta^*=2\pi/N\eta=\pi/\bar u$:
\[\Gamma^*=\{  x(  \ell)=(x(\ell_1),x(\ell_2))|  \ell=(\ell_1,\ell_2)\in\{0,\dots, N-1\}^2\}
,\  x(\ell)=-\bar x+\ell\eta^*,\  \bar x=N\eta^*/2\ .\]

For any $  S_0=e^{  X_0}$ with $  X_0=  x( \ell)\in\Gamma^*$ we then have the approximation
\begin{equation}
\mbox{Spr}(  S_0;T)\sim\frac{\eta^2e^{-rT}}{(2\pi)^2}\sum^{N-1}_{k_1,k_2=0}e^{ i(u( k)+i\epsilon)  x( \ell)'}
\Phi(u(k)+i\epsilon;T)\hat P(  u(  k)+i\epsilon)\ .\end{equation}
Now, as usual for the discrete FFT, as long as $N$ is even, 
\[ iu( k)  x( \ell)'=i\pi(k_1+k_2+\ell_1+\ell_2) +2\pi ik\ell'/N \quad (\mbox{mod}\ 2\pi i)\ .\]
This leads to the double inverse discrete Fourier transform 
\begin{eqnarray}
\mbox{Spr}(  S_0;T)&\sim&(-1)^{\ell_1+\ell_2}e^{-rT}\left(\frac{\eta N}{2\pi}\right)^2e^{-\epsilon x(\ell)' }\left[\frac1{N^2}\sum^{N-1}_{k_1,k_2=0}
e^{2\pi ik \ell '/N}H(  k)\right]\nonumber\\
&=&(-1)^{\ell_1+\ell_2}e^{-rT}\left(\frac{\eta N}{2\pi}\right)^2e^{-\epsilon x(\ell)'}[\mbox{ifft2}(H)](  \ell)
\label{spreadfft}
\end{eqnarray}
where 
\[
H(  k)=(-1)^{k_1+k_2}\Phi(u(k)+i\epsilon;T)\hat P(  u(  k)+i \epsilon)\ .
\]

The selection of suitable values for $\epsilon, N$ and $\eta$ is a somewhat subtle issue when implementing the above FFT approximation of \eqref{spread}. We now briefly discuss separately the truncation error and discretization error. The pure truncation error, defined by taking $\eta\to 0, N\to\infty$ keeping $\bar u=N\eta/2$ fixed, can be made small if the integrand of \eqref{spread} is small outside the square $[-\bar u+i\epsilon_1,\bar u+i\epsilon_1]\times [-\bar u+i\epsilon_2,\bar u+i\epsilon_2]$. Corollary \ref{corol}, proved in the Appendix, gives an $O(|u|^{-2})$ upper bound on $\hat P$, while $\Phi(u)$ can generally been seen directly to have some $u$-decay. Typically, with some caveats, if one picks $\bar u$ large enough so that $|\Phi| <\delta_1\ll 1$ and has decay outside the square, then the truncation error will be less than $O(\delta_1)$.  

The pure discretization error, defined by taking $\bar u\to\infty, N\to\infty$ while keeping $\eta$ fixed, can be made small if $e^{\epsilon X_0'}\mbox{Spr}$, taken as a function of $X_0\in\mathbb{R}^2$, has rapid decay in $X_0$. This is related to the smoothness of $\Phi(u)$ and the choice of $\epsilon$. The first two models are not very sensitive to $\epsilon$, but in the VG model the following conditions are needed to ensure that singularities in $u$ space are avoided:
\[ -a_+<\epsilon_1,\epsilon_2,\epsilon_1+\epsilon_2<a_-\ .
\]
 The error in the approximation \eqref{spreadfft} with $N=\infty$ and $\eta$ finite, is given by the formula
\[\mbox{Spr}^{(\eta)}(X_0)-\mbox{Spr}(X_0)=\sum_{\ell\in\mathbb{Z}^2\backslash \{(0,0)\}} e^{-2\pi\epsilon\ell'/\eta}\mbox{Spr}(X_0+2\ell \bar x)\ .
\]
Typically, with some caveats, the norm of this can be made smaller than $O(\delta_2)\ll 1$ for $X_0\in[-c\bar x,c\bar x]^2$ if $0<c\ll 1$ and $e^{\epsilon X_0'}\mbox{Spr}(X_0)$ is $O(\delta_2)$ and rapidly decaying outside the square $[-\bar x,\bar x]^2$.   

In summary, one expects that the combined truncation and discretization error will be $O(\delta_1+\delta_2)$ if $\bar u$ and $\eta=\pi/\bar x$ are each chosen as above. 

The FFT method can also be applied to the Greeks, enabling us to tackle hedging and other interesting problems. It is particularly efficient for the GBM model, where differentiation under the integral sign is always permissible. For instance, the FFT formula for vega (the sensitivity to $\sigma$) takes the form:\\
\begin{eqnarray*}
\frac{\partial{\mbox{Spr}(  S_0;T)}}{\partial{\sigma_1}}&=&(-1)^{\ell_1+\ell_2}e^{-rT}\left(\frac{\eta N}{2\pi}\right)^2e^{-\epsilon x(\ell)'}[\mbox{ifft2}(\frac{\partial{H}}{\partial{\sigma_1}})](  \ell)\ ;\\
\frac{\partial{H( k)}}{\partial{\sigma_1}}&=&\left[-(u(k)+i\epsilon)\left(i \frac{\partial{\sigma^2}}{\partial{\sigma_1}}'+\frac{\partial{\Sigma}}{\partial{\sigma_1}}(  u(  k)+i \epsilon)'\right)T/2\right] H(k)\ ,
\end{eqnarray*}
where $\frac{\partial{\sigma^2}}{\partial{\sigma_1}}=[2\sigma_1,0]$ and $\frac{\partial{\Sigma}}{\partial{\sigma_1}}=[2\sigma_1,  \rho\sigma_2; \rho\sigma_2, 0]$. Other Greeks including those of higher orders can be computed in similar fashion. This method needs to be used with care for the SV and VG models, since it is possible that differentiation leads to an integrand that decays slowly.  

\section{Numerical Results}
Our numerical experiments were coded and implemented in Matlab version 7.6.0 on an Intel 2.80 GHz machine running under Linux with 1 GB physical memory.  If they were coded in C++ with similar algorithms, we should expect to see much faster performance. All the following experiments were run with $\epsilon_1=-3, \epsilon_2=1$.

 The strength of the FFT method is demonstrated by comparison with benchmarks. Based on a selection of $\{S_{10}^i, S_{20}^i, K^i\}, i\in 1,2,3\dots n$, the objective function we measure is defined as
 \begin{equation}
 \mbox{Err}=\frac{1}{n}\sum_{i=1}^{n}|\log(M^i)-\log(B^i)|
 \label{eq:objfct}
 \end{equation}
 where $M^i$ and $B^i$ are FFT computed prices and benchmark prices for the $ith$ combination. We 
chose $\log{S_{10}}=\frac{i\pi}{10}, \log{S_{20}}=-\frac{\pi}{5}+\frac{j\pi}{10}, i, j\in{1,2,3,4,5,6}$ and fixed the strike $K=1$, leading to 36 prices contributing to the objective function. These choices cover a wide range of moneyness, from deep out-of-the-money to deep in-the-money. Since these combinations all lie on lattices $\Gamma^*$ corresponding to $N=2^n$ and $\bar u/10=2^m$ for integers $n,m$, all 36 prices $M^i$ can be computed simultaneously with one FFT.
  
\begin{figure}[b]
\centering\includegraphics[scale=.7]{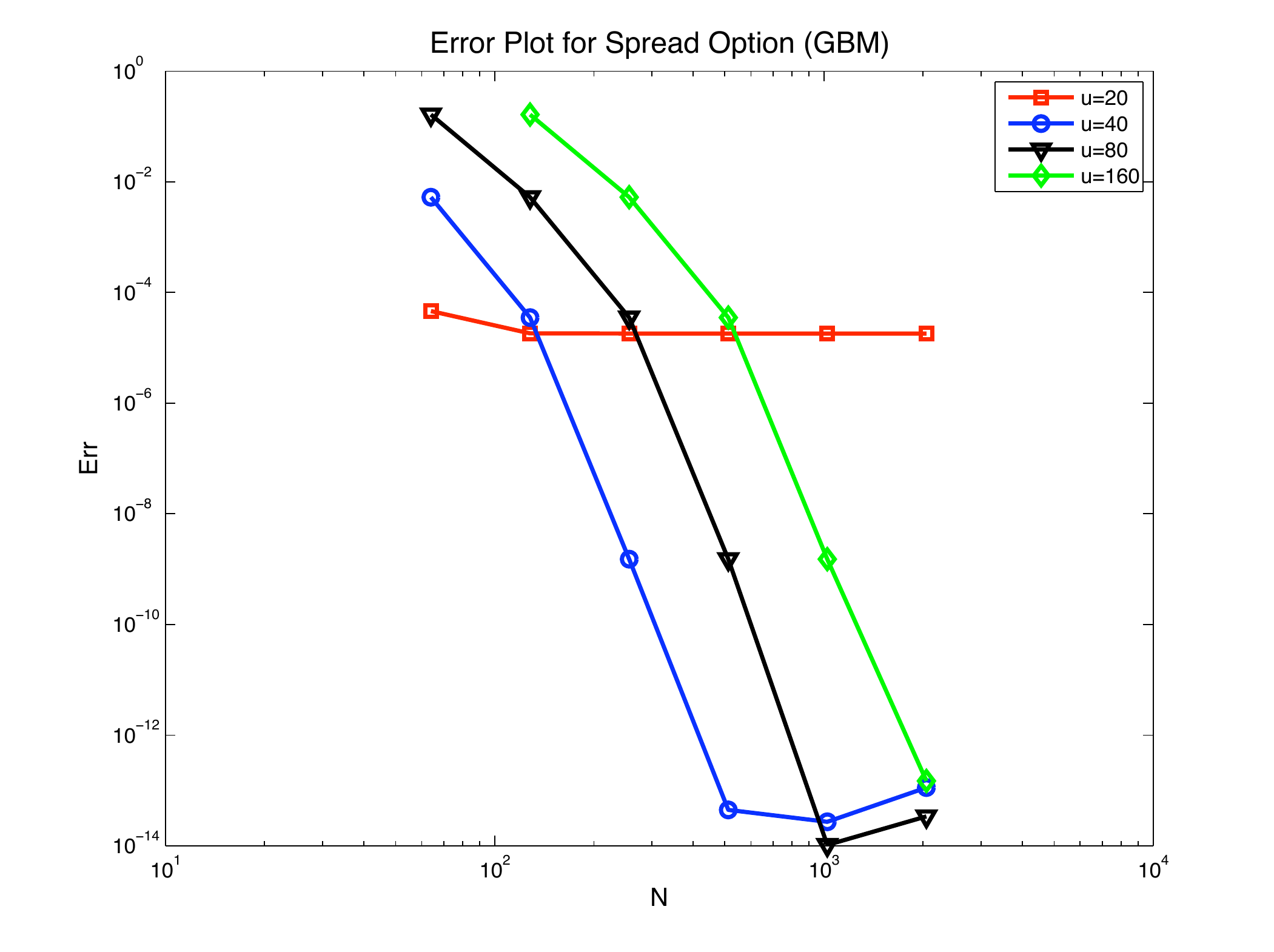}
\caption{{This graph shows the objective function $\rm{Err}$ for the numerical computation of the GBM spread option versus the benchmark. Errors are plotted against the grid size for different choices of $\bar u$. The parameter values are taken from \cite{DempHong00}:
$r = 0.1, T = 1.0, \rho = 0.5, 
\delta_1 = 0.05,\sigma_1 = 0.2,
\delta_2 = 0.05, \sigma_2 = 0.1$. } }
\label{rmserror2dgbm}

\end{figure}

Figure \ref{rmserror2dgbm} shows how the FFT method performs in the 2-dimensional Geometric Brownian motion model for different choices of $N$ and $\bar u$. Since the two factors are bivariate normal, benchmark prices can be calculated to high accuracy by one dimensional integrations. In Figure \ref{rmserror2dgbm} we can clearly see the effects of both truncation errors and discretization errors. For a fixed $\bar u$, the objective function decreases when $N$ increases. The $\bar u=20$ curve flattens out near $10^{-5}$ due to its truncation error of that magnitude. In turn, we can quantify its discretization errors with respect to $N$ by subtracting the truncation error from the total error. The flattening of the other three curves near $10^{-14}$ should be attributed to Matlab round-off errors: because of the rapid decrease of the characteristic function $\Phi$, their truncation error is negligible. For a fixed $N$, increasing $\bar u$ brings two effects: reducing truncation error and enlarging discretization error. These effects are well demonstrated in figure \ref{rmserror2dgbm}.

\begin{figure}[b]
\centering\includegraphics[scale=.7]{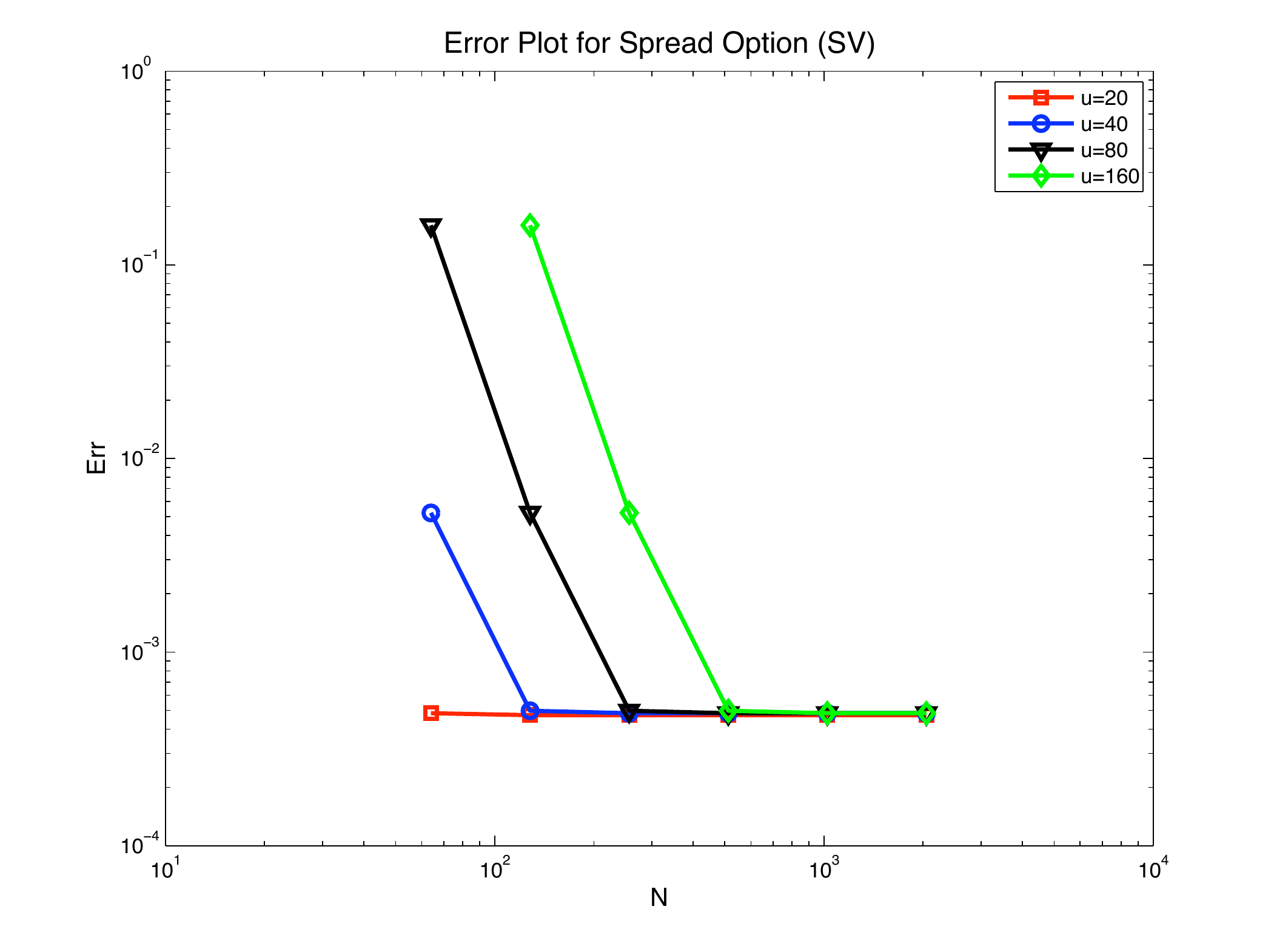}
\caption{{This graph shows the objective function $\rm{Err}$ for the numerical computation of the SV spread option versus the benchmark Monte Carlo simulation values computed with $1,000,000$ simulations each of which consists of 2000 time steps. The parameter values are taken from \cite{DempHong00}:
 $r = 0.1, T = 1.0, \rho = 0.5, 
\delta_1 = 0.05,\sigma_1 = 1.0, \rho_1 = -0.5, 
\delta_2 = 0.05, \sigma_2 = 0.5, \rho_2 =0.25, v_0 = 0.04, \kappa = 1.0, \mu= 0.04, \sigma_v= 0.05$.} }
\label{rmserrorSVsimu}

\end{figure}

For the stochastic volatility model,  no analytical or numerical method we know is consistently more accurate to serve as benchmark. Instead, we used Monte Carlo simulation to achieve a moderate degree of accuracy. Each benchmark price was created by $1,000,000$ simulations, each consisting of 2000 time steps. No variance reduction was employed. The resulting objective function is shown in Figure \ref{rmserrorSVsimu}. The objective function only decreases to near $4\times10^{-4}$, reflecting the inaccuracy of the benchmark. A comparable graph (not shown), using benchmark prices computed with the FFT method with $N=2^{12}$ and $\bar u=80$, showed similar behaviour to Figure \ref{rmserror2dgbm}, and is consistent with accuracies at the level of round-off. The computational cost  to reduce the Monte Carlo simulation error to a comparable level would be prohibitive.

\begin{figure}[b]
\centering\includegraphics[scale=.7]{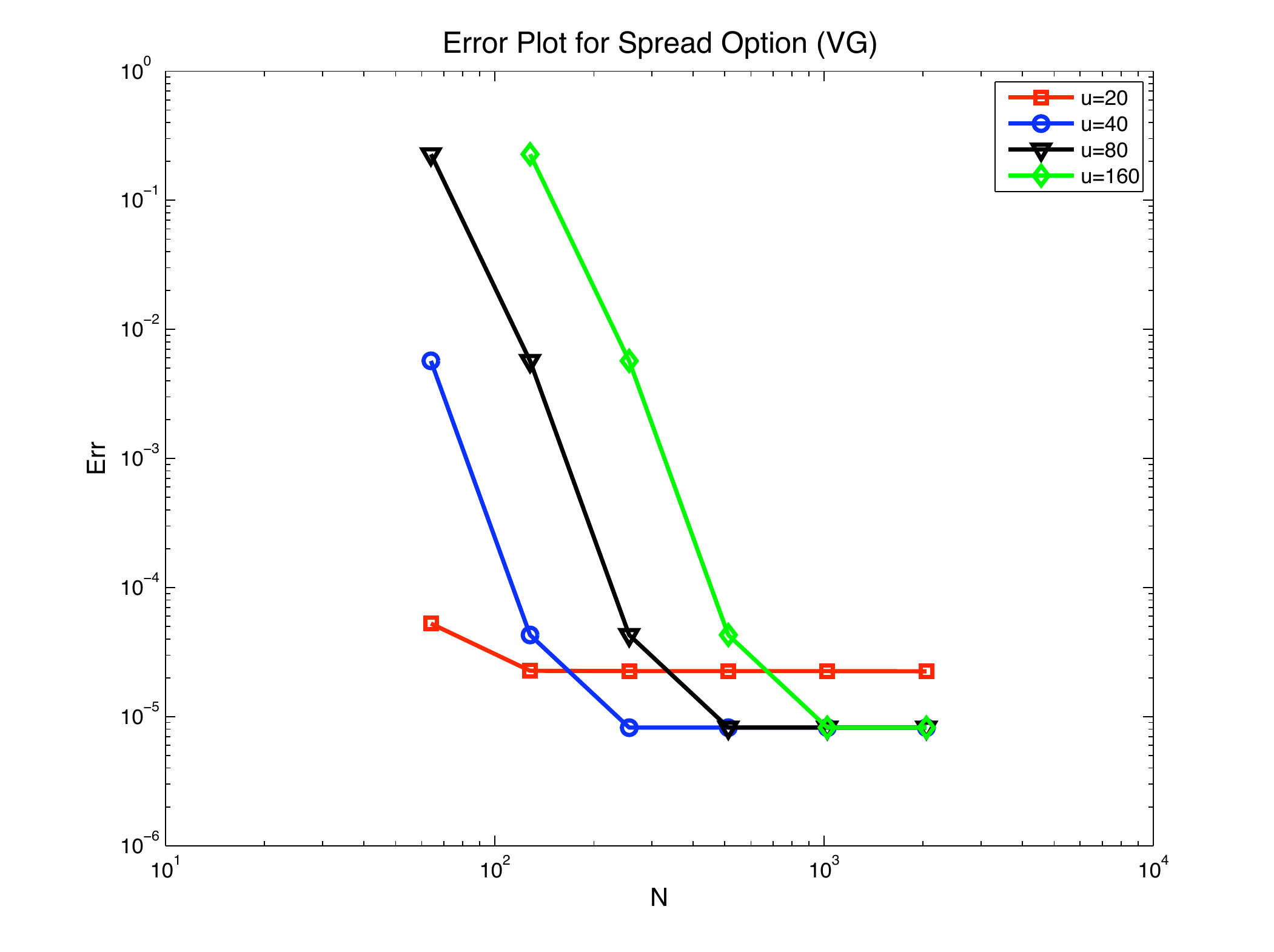}
\caption{{This graph shows  the objective function $\rm{Err}$ for the numerical computation of the VG spread option versus the benchmark values computed with Monte Carlo simulation, where $10^9$ simulations are performed for each price. Errors are plotted against the grid size for five different choices of $\bar u$. The parameters are: $r = 0.1, T = 1.0, \rho = 0.5, a_+=20.4499, a_-=24.4499, \alpha=0.4, \lambda=10$} }
\label{rmserrorVGjump}

\end{figure}

The VG model also lacks a reliable benchmark, so again we used Monte Carlo simulation. Since the VG process can be simulated as a Brownian motion time-changed by a gamma process, MC is here much more efficient than for the  SV model, and we were able to run $10^9$ simulations with antithetic variates for each price. The objective function is shown in Figure \ref{rmserrorVGjump}. The truncation error for $\bar{u}=20$ is about $2\times 10^{-5}$. The other three curves flatten out near $7\times 10^{-6}$, which is presumably the accuracy of the benchmark. A comparable graph (not shown), using benchmark prices computed with the FFT method with $N=2^{12}$ and $\bar u=80$, showed similar behaviour to Figure \ref{rmserror2dgbm}, and is consistent with the FFT method being capable of producing accuracies at the level of round-off.

The strength of the FFT method is further illustrated by the computation of individual prices shown in Tables \ref{tb:twogbm} to \ref{tb:VG}. The columns labeled  ``Analytic'' and ``Simulation'' refer to the type of benchmark used. One can observe that an FFT with $N=256$ is capable of producing very high accuracy in all three models. In the GBM model, the errors are simply due to round-off. The discrepancies in the VG model are small too, less than 1 basis point. The apparent discrepancies in the SV model we attribute to the inaccurate benchmark, rather than the FFT method itself.

\begin{table}[b]
\caption{Prices for the two-factor GBM model of \cite{DempHong00} for different choices of $N$. The parameter values are the same as Figure \ref{rmserror2dgbm} except we fix $S_{10}=100, S_{20}=96, \bar u=40$.}
\begin{center}\footnotesize
\begin{tabular}{|c||c|c|c|c|c|}
\hline
Strike K	&Analytic	&64	&128	&256&	512\\ \hline\hline
0.4 &	8.312461 &	8.206666& 	8.312331 &	8.312461& 	8.312461\\ \hline 
0.8 &	8.114994 &	8.009643& 	8.114864 &	8.114994& 	8.114994\\ \hline 
1.2 &	7.920820 &	7.815913& 	7.920691 &	7.920820& 	7.920820\\ \hline 
1.6 &	7.729932 &	7.625469& 	7.729804 &	7.729932& 	7.729932\\ \hline 
2.0 &	7.542324 &	7.438304& 	7.542196 &	7.542324& 	7.542324\\ \hline
2.4 &	7.357984 &	7.254408& 	7.357857 &	7.357984& 	7.357984\\ \hline 
2.8 &	7.176902 &	7.073770& 	7.176775 &	7.176902& 	7.176902\\ \hline 
3.2 &	6.999065 &	6.896377& 	6.998939 &	6.999065& 	6.999065\\ \hline 
3.6 &	6.824458 &	6.722213& 	6.824332 &	6.824458& 	6.824458\\ \hline 
4.0 &	6.653065 &	6.551264& 	6.652940 &	6.653065& 	6.653065\\ \hline 

\end{tabular}
\end{center}
\label{tb:twogbm}
\end{table}%

\begin{table}[b]
\caption{Prices for the 3 factor SV model of \cite{DempHong00} for different choices of $N$. The parameter values are the same as Figure \ref{rmserrorSVsimu} except we fix $S_{10}=100, S_{20}=96, \bar u=40$. The interpolation is based on a matrix with discretization of $N=256$ and a polynomial with degree of 8. }
\begin{center}\footnotesize
\begin{tabular}{|c||c|c|c|c|c|c|}
\hline
Strike K	&64	&128	&256&	512&	Interpolation&Simulation\\ \hline\hline
2.0 &	6.996467 &	7.544853& 	7.548502 &	7.548502& 	7.548502  &7.545784  \\ \hline
2.2 	&6.902676 &	7.449895 	&7.453536 &	7.453536 	&7.453536& 7.435475  \\ \hline
2.4 	&6.809696 &	7.355748 	&7.359381 &	7.359381 	&7.359381& 7.360184  \\ \hline
2.6 	&6.717527 &	7.262411 	&7.266036 &	7.266037 	&7.266036&  7.276051  \\ \hline
2.8 	&6.626167 &	7.169883 	&7.173501 &	7.173501 	&7.173501& 7.193546 \\ \hline
3.0 	&6.535616 &	7.078165 	&7.081775 &	7.081775 	&7.081775 &7.077421 \\ \hline
3.2 	&6.445873 &	6.987254 	&6.990856 &	6.990857 &	6.990856 &6.983995 \\ \hline
3.4 	&6.356936 &	6.897150 	&6.900745 &	6.900745 &	6.900745&6.913994  \\ \hline
3.6 	&6.268806 &	6.807853 	&6.811439 &	6.811440 &	6.811439&6.818526  \\ \hline
3.8 	&6.181481 &	6.719360 	&6.722939 &	6.722939 &	6.722939&6.735112  \\ \hline
4.0 	&6.094959& 	6.631670 	&6.635241& 	6.635242 &	6.635241&6.621617  \\ \hline
\end{tabular}
\end{center}\label{3fsv}
\end{table}%

\begin{table}[b]
\caption{Prices for the VG model of for different choices of $N$. The parameter values are the same as Figure \ref{rmserrorVGjump} except we fix $S_{10}=100, S_{20}=96, \bar u=40$. The interpolation is based on a matrix with discretization of $N=256$ and a polynomial with degree of 8. For the last column, $10^9$ simulations are used for each price. }
\begin{center}\footnotesize
\begin{tabular}{|c||c|c|c|c|c|c|}
\hline
Strike K	&64	&128	&256&	512&	Interpolation&Simulation\\ \hline\hline
2.0 &	9.157674&	9.723691&	9.727458&	9.727458&	9.727458&9.727546\\ \hline
2.2 &	9.061487&	9.626247&	9.630006&	9.630006&	9.630006&9.629444\\ \hline
2.4 &	8.965876&	9.529448&	9.533200&	9.533200&	9.533200&9.533137\\ \hline
2.6 &	8.870896&	9.433296&	9.437040&	9.437040&	9.437040&9.437323\\ \hline
2.8 &	8.776560&	9.337792&	9.341527&	9.341528&	9.341527&9.341547\\ \hline
3.0 &	8.682870&	9.242934&	9.246662&	9.246662&	9.246662&9.246273\\ \hline
3.2 &	8.589828&	9.148725&	9.152445&	9.152445&	9.152445&9.152956\\ \hline
3.4 &	8.497433&	9.055163&	9.058875&	9.058875&	9.058875&9.058997\\ \hline
3.6 &	8.405687&	8.962250&	8.965954&	8.965954&	8.965954&8.966410\\ \hline
3.8 &	8.314590&	8.869984&	8.873681&	8.873681&	8.873681&8.874129\\ \hline
4.0 &	8.224140&	8.778368&	8.782057&	8.782057&	8.782057&8.781863\\ \hline

\end{tabular}
\end{center}
\label{tb:VG}
\end{table}%

The FFT computes in a single iteration an $N\times N$  panel of prices $spread$ corresponding to initial values $S_{10}=e^{x_{10}+\ell_1\eta^*}$, $S_{20}=e^{x_{20}+\ell_2\eta^*}$, $K=1$, ($\ell_1,\ell_2)\in\{0,\dots, N-1\}^2$. If the desired selection of $\{S_{10}, S_{20}, K\}$ fits into this panel of prices, or its scaling, a single FFT suffices. If not, then one has to match ($x_{10}, x_{20}$) with each combination, and run  several FFTs,  with a consequent increase in computation time. However, we have found that an interpolation technique is very accurate for practical purposes. For instance, prices for multiple strikes with the same $S_{10}$ and $S_{20}$ are approximated by a polynomial fit along the diagonal of the price panel: $\mbox{Spr}(S_0;K_1)=K_1\cdot spread(1,1)$, $\mbox{Spr}(S_0;K_1e^{-\eta^*})=K_1e^{-\eta^*}\cdot spread(2,2)$, $\mbox{Spr}(S_0;K_1e^{-2\eta^*})=K_1e^{-2\eta^*}\cdot spread(3,3)\dots$. 
The results of this technique are recorded in Table \ref{3fsv} and Table \ref{tb:VG} in the column ``Interpolation''. We can see this technique generates very competitive results and moreover, saves computational resource.

Finally, we computed first order Greeks using the method described at the beginning of Section \ref{nifft} and compared them with finite differences. As seen in Table \ref{tb:greeks}, the two methods come up with very consistent results. The Greeks of our at-the-money spread option exhibit some resemblance to the at-the-money European put/call option. The delta of $S_1$ is close to the delta of the call option, which is about 0.5. And the delta of $S_2$ is close to the delta of the put option, which is also about 0.5. The time premium of the spread option is positive. The option price is much more sensitive to $S_1$ volatility than to $S_2$ volatility. It is an important feature that the option price is negatively correlated with the underlying correlation: Intuitively speaking, if the two underlyings are strongly correlated, their co-movements diminish the probability that $S_{1T}$ develops a wide spread over $S_{2T}$. This result is consistent with observations made by \cite{DenLiZho06}.

\begin{table}[b]
\caption{The Greeks for the GBM model compared between the FFT method and the finite difference method. The FFT method uses  $N=2^{10}$ and $\bar u=40$. The finite difference uses a two-point central formula, in which the displacement is $\pm 1\%$. Other parameters are the same as Table \ref{tb:twogbm} except that  we fix the strike $K=4.0$ to make the option at-the-money.}
\begin{center}\footnotesize
\begin{tabular}{|c||c|c|c|c|c|c|}
\hline
	&Delta(S1)&Delta(S2)&Theta&Vega($\sigma_1$)&Vega($\sigma_2$)&$\partial{\mbox{Spr}}/\partial{\rho}$\\ \hline\hline
FD &0.512648 &-0.447127& 3.023823 &33.114315& -0.798959  & -4.193749 \\ \hline
FFT &0.512705 &	 -0.447079& 3.023777& 33.114834	& -0.798972& -4.193728  \\ \hline

\end{tabular}
\end{center}
\label{tb:greeks}
\end{table}%

Since the FFT method naturally generates a panel of prices, and interpolation can be implemented accurately with negligible additional computational cost,  it is appropriate to  measure the efficiency of the method by timing the computation of a panel of prices.  Such computing times are shown in Table \ref{tb:ffttime}. For the FFT method, the main computational cost comes from the calculation of the matrix $H$ in \eqref{spreadfft} and the subsequent FFT of $H$. We see that the GBM model is noticeably faster than the SV and VG models: This is due to a recursive method used to calculate the $H$ matrix entries of the GBM model,  which is not applicable for the SV and VG models. The number of calculations for $H$ is of order$N^2$ which for large $N$ exceeds the $N\log{N}$ of the FFT of $H$, and thus the advantage of this efficient algorithm for GBM is magnified as $N$ increases. However, our FFT method is still very fast for the SV and VG models and is able to generate a large panel of prices within a couple of seconds.

\begin{table}[b]
\caption{Computing time of FFT for a panel of prices.}
\begin{center}\footnotesize
\begin{tabular}{|c||c|c|c|}
\hline
Discretization&	GBM&	SV&	VG\\ \hline\hline
64&	0.091647&	0.083326&	0.109537\\ \hline
128&	0.099994&	0.120412&	0.139276\\ \hline
256&	0.126687&	0.234024&	0.220364\\ \hline
512&	0.240938&	0.711395&	0.621074\\ \hline
1024&	0.609860&	2.628901&	2.208770\\ \hline
2048&	2.261325&	10.243228&	8.695122\\ \hline
\end{tabular}
\end{center}
\label{tb:ffttime}
\end{table}%

\section{High Dimensional Spread Options}
The ideas of section 2 turn out to extend naturally to basket options with payoffs $(\tilde S_T-S_{1T}-\dots-S_{MT}-1)^+$ for $M\ge 2$. The analysis is based on a simple result proved in the Appendix:
\begin{proposition} \label{prop2}
Let  $z\in \mathbb{R}$ and $u=(u_1,\dots,u_M)'\in\mathbb{C}^M$ with $\Im m( u_m)>0$ for all $m\le M$. Then
\begin{equation}\int_{\mathbb{R}^M}\ e^z\delta(e^z-\sum_{m=1}^M e^{x_m}) e^{-iux'} d^Mx=\frac{\prod_{m=1}^M\Gamma(-iu_m)}{\Gamma(-i\sum_{m=1}^M u_m)} e^{-i(\sum_{m=1}^M u_m)z}\ .
\end{equation}
\end{proposition}

As before we write $x_m=\log S_m, \tilde x=\log\tilde S$ and seek to compute for $u\in \mathbb{C}^M$, $\tilde u\in\mathbb{C}$
\[\hat P(u,\tilde u)=\int_{\mathbb{R}^{M+1}}(e^{\tilde x}-\sum_{m=1}^M e^{x_m}-1)^+ e^{-i\tilde u\tilde x} e^{-iu'x} d^{M}x d\tilde x\]
We introduce the factor $1=\int_{\mathbb{R}}\delta(e^z-\sum_{m=1}^M e^{x_m}) e^z dz$ and interchange the $z$ integral with the $x$ integrals. Then using Proposition \ref{prop2} one finds
\begin{eqnarray*} 
\hat P(u,\tilde u)&=&\int_{\mathbb{R}^{2}}e^{-i\tilde u\tilde x}(e^{\tilde x}-\sum_{m=1}^M e^{x_m}-1)^+\left[\int_{\mathbb{R}^M}e^z\delta(e^z-\sum_{m=1}^M e^{x_m}) e^{-iux'} d^{M}x\right]d\tilde xdz\\
&=&\int_{\mathbb{R}^{2}}e^{-i\tilde u\tilde x}(e^{\tilde x}-e^z-1)^+\left[\int_{\mathbb{R}^M}e^z\delta(e^z-\sum_{m=1}^M e^{x_m}) e^{-iux'} d^{M}x\right]d\tilde xdz\\
&=&\frac{\prod_{m=1}^M\Gamma(-iu_m)}{\Gamma(-i\sum_{m=1}^M u_m)} \int_{\mathbb{R}^{2}}e^{-i\tilde u\tilde x}e^{-i(\sum_{m=1}^M u_m)z}(e^{\tilde x}-e^z-1)^+d\tilde xdz
\end{eqnarray*}
When we can apply Proposition 1, we  obtain the result:
\begin{proposition} \label{prop3}
For any real numbers $\tilde \epsilon, \epsilon=(\epsilon_1,\dots, \epsilon_M)$ with $\epsilon_m>0$ for all $m\le M$ and $\tilde \epsilon\le -1-\sum_{m=1}^M \epsilon_m$
\begin{equation}
(e^{\tilde x}-\sum_{m=1}^M e^{x_m}-1)^+=(2\pi)^{-(M+1)}\int_{\mathbb{R}^{M+1}+i(\epsilon,\tilde\epsilon)} e^{i\tilde u\tilde x+iux'} \hat P(u,\tilde u) d^Mu \ d\tilde u
\label{pfftM}
 \end{equation}
where
\begin{equation}\label{multispread} \hat P(u,\tilde u)=\frac{\Gamma(i(\tilde u+\sum_{m=1}^M u_m)-1)\prod_{m=1}^M\Gamma(-iu_m)}{\Gamma(i\tilde u+1)}.
\end{equation}

\end{proposition}

\section{Conclusion} This paper presents a fundamentally new approach to the valuation of spread options, an important class of financial contracts. The method is based on a newly discovered explicit formula for the Fourier transform  of the spread option payoff in terms of the gamma function. 

This mathematical result leads to simple and transparent algorithms for computing spread options in all dimensions. The powerful tool of the Fast Fourier Transform  provides an accurate and efficient implementation of the fundamental result. The difficulties and pitfalls of the FFT, of which there are admittedly several, are by now well understood, and thus the reliability and stability properties of our method are rather clear. 

Many important processes in finance, particularly affine models and L\'evy jump models, have well known explicit characteristic functions, and can be included in the method with little difficulty. Thus the method can be easily applied to important problems arising in energy and commodity markets. 

Finally, the Greeks can be systematically evaluated for such models, with similar performance and little extra work.

\begin{appendix}
\section{Proof of Theorem \ref{mainprop} and Proposition \ref{prop2}}

\noindent{\bf Proof of Theorem \ref{mainprop}:\ }
Suppose  $\epsilon_2>0,\epsilon_1+\epsilon_2<-1$. One can then verify either directly or from the argument that follows that $e^{  \epsilon\cdot  x}P(  x),  \epsilon=(\epsilon_1,\epsilon_2)$ is in $\mathbb{L}^2(\mathbb{R}^2)$. Therefore, application of the Fourier inversion theorem to $e^{  \epsilon\cdot  x}P(  x),  \epsilon=(\epsilon_1,\epsilon_2)$ implies that 
\be P(  x)=(2\pi)^{-2}\iint_{\mathbb{R}^2+i\epsilon}e^{i  u\cdot  x}g(  u) d^2  u
\label{pft}\ee
where 
\[ g(  u)=\iint_{\mathbb{R}^2}e^{-i  u\cdot  x}P(  x) d^2  x\ .\]
By restricting to the domain $\{  x: x_1>0, e^{x_2}<e^{x_1}-1\}$ we have
\beqq
g(  u)&=& \int^\infty_0e^{-iu_1x_1}\left[\int^{\log(e^{x_1}-1)}_{-\infty} e^{-iu_2x_2}[(e^{x_1}-1)-e^{x_2}]dx_2\right]dx_1
\\
&=& \int^\infty_0e^{-iu_1x_1}(e^{x_1}-1)^{1-iu_2}\left[\frac1{-iu_2}-\frac1{1-iu_2}\right]dx_1\ .
\eeqq
The change of variables $z=e^{-x_1}$ then leads to
\[g(  u)=\frac1{(1-iu_2)(-iu_2)}\int^1_{0}z^{iu_1}\left(\frac{1-z}{z}\right)^{1-iu_2}\frac{dz}{z}\ .\]
The beta function 
\[B(a,b)=\frac{\Gamma(a)\Gamma(b)}{\Gamma(a+b)}\]
is defined for any complex $a,b$ with $\Re e( a),\Re e(b)>0$ by
\[B(a,b)=\int_0^1 z^{a-1}(1-z)^{b-1} dz\ .\]
From this, and the property $\Gamma(z)=(z-1)\Gamma(z-1)$ follow the formulas
\be\label{gform}
g(  u)=\frac{\Gamma(i(u_1+u_2)-1)\Gamma(-iu_2+2)}{(1-iu_2)(-iu_2)\Gamma(iu_1+1)}=\frac{\Gamma(i(u_1+u_2)-1)\Gamma(-iu_2)}{\Gamma(iu_1+1)}\ .
\ee

\qed

The above derivation also leads to the following bound on $\hat P$.
\begin{corollary}\label{corol} Fix $\epsilon_2=\epsilon, \epsilon_1=-1-2\epsilon$ for some $\epsilon>0$. Then
\be\label{phatbound}
|\hat P(u_1,u_2)|\le \frac{\Gamma(\epsilon)\Gamma(2+\epsilon)}{\Gamma(2+2\epsilon)}\cdot\frac1{Q(\|u\|^2/5)^{1/2}}
\ee
where $Q(z)=(z+\epsilon^2)(z+(1+\epsilon)^2)$.
\end{corollary}

\noindent
{\bf Proof: \ } First note that for $z_1,z_2\in\mathbb{C}$, $|B(z_1,z_2)|\le B(\Re e(z_1),\Re e(z_2))$. Then \eqref{gform} and a symmetric formula with $u_2\leftrightarrow -1-u_1-u_2$ leads to the upper bound
\[ \|\hat P(u_1-i(\epsilon+1),u_2+i\epsilon)\|\le B(\epsilon,2+\epsilon)\min\left(\frac1{Q(|u_2|)},\frac1{Q(|u_1+u_2|)}\right)\ .
\]
But since $Q$ is monotonic and $\|u\|\le \sqrt{5}\max\left(|u_2|,|u_1+u_2|\right)$ for all $u\in\mathbb{R}^2$, the required result follows. \qed

\noindent{\bf Proof of Proposition \ref{prop2}:\ }
To simplify the proof, we make the change of variables $p=e^z$ and $q_m=e^{x_m}$. It becomes to prove that
\beq
\int_{\mathbb{R}^M}\ p\delta(p-\sum_{m=1}^M q_m)\prod_{m=1}^M q_{m}^{-iu_m-1} d^Mq=\frac{\prod_{m=1}^M\Gamma(-iu_m)}{\Gamma(-i\sum_{m=1}^M u_m)} p^{-i(\sum_{m=1}^M u_m)}\ .
\eeq

We prove the proposition by induction. The above equation trivially holds when $M=1$. Now suppose it holds for $M=N$. Then for $M=N+1$ 
\beq 
LHS&=&\int_{\mathbb{R}^{N+1}}\ p\delta(p-q_{N+1}-\sum_{m=1}^N q_m)q_{N+1}^{-iu_{N+1}-1}\prod_{m=1}^N q_{m}^{-iu_m-1} d^{N+1}q\nonumber \\
&=&\frac{\prod_{m=1}^N\Gamma(-iu_m)}{\Gamma(-i\sum_{m=1}^N u_m)} \int_0^p \frac{p}{p-q_{N+1}}(p-q_{N+1})^{-i(\sum_{m=1}^N u_m)} q_{N+1}^{-iu_{N+1}-1} dq_{N+1}\ .
\eeq
The above step makes use of the result when $M=N$. The proof is then complete when one notices that the $q_{N+1}$ integral is simply $p^{-i(\sum_{m=1}^{N+1} u_m)}$ multiplied by a beta function with parameters $-i(\sum_{m=1}^N u_m)$ and $-iu_{N+1}$.

\qed
 
\end{appendix}

\bibliographystyle{siam}


\end{document}